\documentclass[runningheads]{llncs}
\usepackage[utf8]{inputenc} % allow utf-8 input
\usepackage[T1]{fontenc}    % use 8-bit T1 fonts
\usepackage{hyperref}       % hyperlinks
\usepackage{url}            % simple URL typesetting
\usepackage{booktabs}       % professional-quality tables
   
\usepackage{amsfonts}       % blackboard math symbols
\usepackage{nicefrac}       % compact symbols for 1/2, etc.
\usepackage{microtype}      % microtypography
\usepackage{cite}
\usepackage{amsmath,amssymb}
\usepackage{algorithm, algorithmic}
\usepackage{url}
\usepackage{pgf}
\usepackage{tikz}
  \usetikzlibrary{arrows,automata}
  \usetikzlibrary{spy}
\usepackage{pgfplots}
  \pgfplotsset{width=7cm,compat=1.8}
\usepackage{multirow}
\usepackage{tabulary}
\usepackage{enumitem}
\newcommand{\Op}[1]{\operatorname{\mathcal{#1}}}

\newcommand{\x}{\boldsymbol{x}}
\newcommand{\y}{\boldsymbol{y}}

\newcommand{\e}{\boldsymbol{e}}
\newcommand{\h}{\boldsymbol{h}}

\newcommand{\X}{\mathbb{X}}
\newcommand{\Y}{\mathbb{Y}}
\newcommand{\R}{\mathcal{R}}
\newcommand{\G}{\mathcal{G}}
\newcommand{\W}{\mathcal{W}}
\newcommand{\D}{\mathcal{D}}

\newtheorem{prop}{Proposition}

\begin{document}
%
%\title{Contribution Title\thanks{Supported by organization x.}}

\title{Adversarially learned iterative reconstruction for imaging inverse problems}


\author{Subhadip Mukherjee$^1$ \and Ozan \"Oktem$^2$ \and Carola-Bibiane Sch\"onlieb$^1$}

\institute{Department of Applied Mathematics and Theoretical Physics\\ University of Cambridge, UK\\
\email{\{sm2467, cbs31\}@cam.ac.uk}\\
\and
Department of Mathematics, KTH -- Royal Institute of Technology, Sweden\\
\email{ozan@kth.se}}
\maketitle              % typeset the header of the contribution
\begin{abstract}
In numerous practical applications, especially in medical image reconstruction, it is often infeasible to obtain a large ensemble of ground-truth/measurement pairs for supervised learning. Therefore, it is imperative to develop unsupervised learning protocols that are competitive with supervised approaches in performance. Motivated by the maximum-likelihood principle, we propose an unsupervised learning framework for solving ill-posed inverse problems. Instead of seeking pixel-wise proximity between the reconstructed and the ground-truth images, the proposed approach learns an iterative reconstruction network whose output matches the ground-truth in distribution. Considering tomographic reconstruction as an application, we demonstrate that the proposed unsupervised approach not only performs on par with its supervised variant in terms of objective quality measures, but also successfully circumvents the issue of over-smoothing that supervised approaches tend to suffer from. The improvement in reconstruction quality comes at the expense of higher training complexity, but, once trained, the reconstruction time remains the same as its supervised counterpart.     
%to encourage the reconstruction to match the ground-truth in distribution. 
% (such as the peak signal to noise ratio (PSNR) and the structural similarity index (SSIM))

\keywords{Generative adversarial networks (GANs)  \and Iterative reconstruction \and Unsupervised learning \and Inverse problems}
\end{abstract}

\section{Introduction}
Inverse problems are encountered in a wide range of scientific and engineering applications, especially in the context of medical imaging. The aim is to recover an unknown model parameter $\x^*\in\X$ containing critical information about the structural details of an underlying subject from data $\y = \Op{A}(\x^*)+ \e \in \Y$, representing noisy, indirect, and potentially incomplete set of measurements. The forward operator $\Op{A}:\X\rightarrow \Y$ and the distribution of the measurement noise $\e\in \Y$ are typically known and they jointly form a simulator for the data acquisition process. Generally, $\X$ is the Hilbert space of functions defined on some $\Omega\subseteq \mathbb{R}^d$, while $\Y$ is another Hilbert space of functions defined on a suitable data manifold $\mathbb{M}$. Inverse problems are typically ill-posed in the absence of any further information apart from the measurements alone, meaning that different model parameters can give rise to the same measurement. Variational reconstruction \cite[Part II]{scherzer2009variational} is a generic, yet adaptable framework for solving inverse problems, wherein ill-posedness is tackled by solving 
\begin{equation}
    \underset{\x \in \X}{\min}\text{\,\,}\left\|\y-\Op{A}(\x)\right\|_2^2+\lambda\, \R(\x).
    \label{var_recon}
\end{equation}
The goal here is to alleviate the aforementioned inherent indeterminacy by incorporating some prior knowledge about the model parameter using a hand-crafted regularizer $\R$ in addition to seeking data-consistency. The role of the regularizer is to penalize unlikely or undesirable solutions. The variational framework is said to be well-posed if it admits a unique solution which varies continuously in the measurement. 

While variational methods enjoy rigorous theoretical guarantees for stability and convergence, and have remained the state-of-the-art for several decades, they are limited in their ability to adapt to a particular application at hand. With the emergence of deep learning, modern approaches for solving inverse problems have increasingly shifted towards data-driven reconstruction \cite{data_driven_inv_prob}, which generally offers significantly superior reconstruction quality as compared to the traditional variational methods. Data-adaptive reconstruction methods can broadly be classified into two categories: (i) end-to-end trained over-parametrized models that either attempt to map the measured data to the true model parameter (such as AUTOMAP proposed in \cite{automap}), or remove artifacts from the output of an analytical reconstruction method \cite{postprocessing_cnn}, and (ii) learning the image prior using a neural network based on training data of images and then using such a learned regularizer in a variational model for reconstruction \cite{ar_nips,nett_paper,kobler2020total,meinhardt2017learning}. The first approach relies on learning the reconstruction method from a large training dataset that consists of many ordered pairs of model-parameter and corresponding noisy data. Since obtaining a vast amount of paired examples is difficult in medical imaging applications, over-parametrized models trained end-to-end in a supervised manner might run into the danger of overfitting and generalize poorly on unseen data. The second category still requires one to solve a high-dimensional variational problem where the objective involves a trained neural network, a task that is typically computationally demanding.

% The generalizability of fully data-driven models with many parameters largely depends on the amount of data that they are trained on. As the underlying data get more diverse, such models require more paired examples to learn in order to generalize well on novel samples. Since obtaining a vast amount of paired examples is difficult in medical imaging applications, over-parametrized models trained end-to-end in a supervised manner might run into the danger of overfitting and generalize poorly on new data. 
% \CBS{need references here for fully learned approaches, e.g. AUTOMAP.}
\indent One promising way to circumvent the limited data problem is to build network architectures by incorporating the physics of the acquisition process \cite{lpd_tmi,jonas_learned_iterative}. The learned primal-dual (LPD) approach proposed in \cite{lpd_tmi} is data-efficient as compared to fully data-driven approaches and can generalize well when trained on a moderate amount of examples. However, an unrolled LPD network trained by minimizing the squared-$\ell_2$ error between the network output and the target essentially returns an approximation to the conditional-mean estimator, which is the statistical expectation of the target image conditioned on the measurement. Owing to this implicit averaging, the resulting estimate tends to suffer from blurring artifacts with the loss of important details in the reconstruction. The proposed adversarially trained LPD method, referred to as ALPD, circumvents this problem by seeking proximity in the space of distribution instead of aiming to minimize the squared-$\ell_2$ distortion in the image space. 

\section{Main Contributions: training objective and protocol}
For supervised learning, one needs paired training examples of the form $\left\{\x_i,\y_i\right\}_{i=1}^{n}$ sampled i.i.d. from the joint probability distribution $\pi_X(\x)\pi_{\text{data}}(\y|\x)$ of the image and the measurement. In contrast, the proposed training protocol is unsupervised, i.e., it assumes availability of i.i.d. samples $\left\{\x_i\right\}_{i=1}^{n_1}$ and $\left\{\y_i\right\}_{i=1}^{n_2}$ from the marginal distributions $\pi_X$ and $\pi_Y$ of the ground-truth image and measurement data, respectively. The image and the data samples are unpaired, i.e., $\y_i$ does not necessarily correspond to the noisy measurement of $\x_i$. In the context of CT, $\x_i$'s could be the high-/normal-dose reconstructions obtained using the classical filtered back-projection algorithm, whereas $\y_i$'s correspond to low-dose projection. 
% In medical imaging applications such as CT, the unsupervised learning paradigm is more realistic since it is relatively easier to obtain unpaired data for training. For example, one can, in principle, make use of previously computed high-dose reconstructions of anthropomorphic phantoms while learning to reconstruct from low-dose projection data acquired from human subjects, without having to acquire high-dose projections from them.    
% the first one of which is motivated by supervised learning and the other two are rooted in the philosophy that the variational framework relies on.
% is to learn $\G_{\theta}$ so that it is approximately a left-inverse of $\Op{A}$, i.e., the $\G_{\theta}$ can recover the ground-truth from noise-free measurement.    
%\section{Adversarially learned iterative reconstruction}
We begin with a description of the proposed unsupervised approach and how it differs from and relates to supervised and classical variational methods. Subsequently, we motivate the training loss using the maximum-likelihood (ML) principle and explain the reconstruction network parametrization, which follows the same philosophy proposed in \cite{lpd_tmi}.
\subsection{Proposed training protocol for ALPD}
\label{sec:adv_learning_proposed}
Similar to supervised training, one key component of the proposed unsupervised approach is to first build a parametric reconstruction network $\G_{\theta}:\Y\rightarrow \X$ (see Sec. \ref{g_phi_pdhg_sec} for details) that takes the measurement as input and produces a reconstructed image as the output. However, unlike supervised training, it is not possible to train $\G_{\theta}$ by minimizing a chosen distortion measure between $\G_{\theta}(\y_i)$ and $\x_i$, since $\x_i$ is not the ground-truth image corresponding to $\y_i$. Our training framework essentially seeks to achieve the following three objectives:
\begin{enumerate}[leftmargin=*]
\item The reconstructions produced by $\G_{\theta}$ should be close to the ground-truth images in the training dataset in terms of distribution (measured with respect to the Wasserstein distance); 
\item $\G_{\theta}$ should be encouraged to be the right-inverse of $\Op{A}$, so that the forward operator applied on the output of $\G_{\theta}$ is close to the measured data; and
\item $\G_{\theta}$ should approximately be a left-inverse of $\Op{A}$, i.e., $\G_{\theta}$ must recover the ground-truth from noise-free measurement.
\end{enumerate}
More concretely, we propose to learn $\G_{\theta}:\Y\rightarrow \X$ by minimizing the training loss
\begin{multline}\label{alpd_train_loss}
    J(\theta) = \W\left((\G_{\theta})_{\#}\pi_{Y},\pi_X\right)+\lambda_{\Y}\,\,\mathbb{E}_{\pi_{Y}}\left[\left\|\Op{A}(\G_{\theta}(Y))-Y \right\|_2^2\right]
    \\ + \lambda_{\X}\,\,\mathbb{E}_{\pi_{X}}\left[\left\|\G_{\theta}\left(\Op{A}(X)\right)-X \right\|_2^2\right].
\end{multline}
The penalty parameters $\lambda_{\X}$ and $\lambda_{\Y}$ control the relative weighting of the three objectives. Notably, in the absence of noise in the measurement, the first objective becomes superfluous, i.e., any $\G_{\theta}$ that satisfies the third objective automatically satisfies the first one too. For noisy measurements, the combination of the first and the third objectives helps compute a stable estimate which does not overfit to noise, whereas the second objective ensures that the reconstruction explains the data well. Similar to \cite{wgan_main}, we make use of the Kantorovich-Rubinstein (KR) duality for approximating the Wasserstein distance term in \eqref{alpd_train_loss}. This requires training a critic network $\D_{\alpha}:\X \rightarrow{\mathbb{R}}$ that scores an image on the real line based on how closely it resembles the ground-truth images in the dataset. More precisely, the KR duality helps estimate the Wasserstein distance by solving
% \CBS{Do we need to motivate why we are using the Wasserstein distance for the first objective?}     
% While the first objective is tantamount to imposing the prior knowledge on the solution by leveraging the true image distribution $\pi_X$ available in the form of samples during training, the second objective is akin to seeking data-consistency in the variational reconstruction framework. The third objective, on the other hand, aims to emulate the loss in supervised training by simulating the noise-free data from the image samples while utilizing the knowledge of the forward operator. 
\begin{equation}
\W\left((\G_{\theta})_{\#}\pi_{Y},\pi_X\right)
=\sup_{\alpha}
\mathbb{E}_{\pi_X}\left[\D_{\alpha}\left(X\right)\right]-\mathbb{E}_{(\G_{\theta})_{\#}\pi_{\Y}}\left[\D_{\alpha}\left(X\right)\right]
\text{ where $\D_{\alpha}\in \mathbb{L}_1$.}
\label{wasserstein_dist}
\end{equation}
Here, $\mathbb{L}_1$ denotes the space of 1-Lipschitz functions. In practice, both $\G_{\theta}$ and $\D_{\alpha}$ are updated in an alternating manner instead of fully solving \eqref{wasserstein_dist} for each $\G_{\theta}$ update. The 1-Lipschitz condition is enforced by penalizing the gradient of the critic with respect to the input \cite{wgan_gp}. Estimating the Wasserstein distance in \eqref{wasserstein_dist} and the training loss for $\G_{\theta}$ in \eqref{alpd_train_loss} requires samples from the marginals, thereby rendering the training framework unsupervised. The detailed steps involved in training the networks are listed in Algorithm \ref{algo_acr_train}.

\indent At this point, it is instructive to interpret the training objective \eqref{alpd_train_loss} through the lens of the variational framework, by recasting the variational problem as a minimization over the parameter $\theta$ of $\G_{\theta}$ instead of $\x$. Given the data distribution $\pi_Y$, it is natural to estimate $\theta$ that minimizes the expected variational loss:
\begin{equation}
    \underset{\theta}{\min}\,\,J_1(\theta) := \mathbb{E}_{\pi_Y}\left[\left\|\Op{A}(\G_{\theta}(Y))-Y \right\|_2^2+\lambda\,\R\left(\G_{\theta}(Y)\right)\right].
    \label{var_opt_interpret}
\end{equation}
Now, suppose the existence of an \textit{ideal} regularizer in \eqref{var_opt_interpret}, which returns a small score when the input is drawn from $\pi_X$ and a large score when the distribution of the input differs from $\pi_X$. For such a regularizer, the difference
\begin{equation}
    \mathcal{L}(\R)=\mathbb{E}_{\pi_X}\left[\R(X)\right] - \mathbb{E}_{\pi_Y}\left[\R(\G_{\theta}(Y))\right],
    \label{var_opt_interpret1}
\end{equation}
should be small. As a matter of fact, a consequence of the KR duality is that 
\[
\inf_{\R} \mathcal{L}(\R)
= -\W\left((\G_{\theta})_{\#}\pi_{Y},\pi_X\right),
\]
provided that $\R$ is constrained to be 1-Lipschitz. Substituting this in \eqref{var_opt_interpret1} and ignoring terms independent of $\theta$ reduces \eqref{var_opt_interpret} to minimizing $\theta \mapsto \hat{J}_1(\theta)$ where
\begin{equation}
    \hat{J}_1(\theta) := \mathbb{E}_{\pi_Y}\left[\left\|\Op{A}(\G_{\theta}(Y))-Y \right\|_2^2\right]+\lambda\,\W\left((\G_{\theta})_{\#}\pi_{Y},\pi_X\right).
    \label{var_opt_interpret_red}
\end{equation}
If $\lambda:=1/\lambda_{\Y}$, then we obtain the inequality $J(\theta)\geq \lambda_{\Y}\, \hat{J}_1(\theta)$ for the training loss $J$. This indicates that our training loss majorizes (up to a scaling) the objective $\hat{J}_1$ in \eqref{var_opt_interpret_red}, which emerges naturally from the variational loss under the assumption of a 1-Lipschitz ideal regularizer.
% :
% \begin{equation}
%     J(\theta)\geq \lambda_{\Y}\, \hat{J}_1(\theta).
%     \label{mm_train_loss}
% \end{equation}
The penalty terms in the $\X$- and $\Y$-domains in \eqref{alpd_train_loss} are unmistakably reminiscent of the cycle-consistency losses in cycle-GANs \cite{cycle_gan_main} that are widely used learning paradigms for unpaired image-to-image translation problems. Similar unsupervised approaches involving GANs were also proposed in \cite{conditional_im2im_cvpr} for conditional image-to-image synthesis tasks. The proposed approach can indeed be thought of as a simpler variant of cycle-GAN with the generator learned in only one direction ($\Y\rightarrow \X$) instead of two. Notably, it was recently shown in \cite{sim_cyclegan} that the cycle-GAN training loss can be derived as the optimal transport loss corresponding to the case where the transport cost is equal to the variational loss with $\R(\x)=\left\|\x-\G_{\theta}(\y)\right\|_2$.

\subsection{A maximum-likelihood (ML) perspective}
The ML principle seeks to solve
\begin{equation}
   \underset{\theta}{\max}\,\, \left[\frac{1}{n_1}\sum_{i=1}^{n_1}\log \pi^{(\theta)}_X(\x_i)+\frac{1}{n_2}\sum_{i=1}^{n_2}\log \pi^{(\theta)}_Y(\y_i)\right],
    \label{overall_ml}
\end{equation}
where $\pi^{(\theta)}_X$ and $\pi^{(\theta)}_Y$ are the distributions induced by appropriately postulated probabilistic models on $X$ and $Y$, respectively. In the following, we explain the statistical models and use them to derive a tractable lower-bound (generally referred to as the evidence lower-bound (ELBO)) on the ML objective in \eqref{overall_ml}. Our analysis reveals that the resulting ELBO is equivalent to the ALPD training loss in \eqref{alpd_train_loss} in spirit, except for the measure of distance for comparing the distributions of the reconstruction and the ground-truth. The KL-divergence-based distance measure is replaced with the Wasserstein-1 distance since it lends itself to continuous differentiability with respect to the parameters of the reconstruction network, thereby facilitating a stable gradient-based parameter update.   
\subsubsection{Bound on the data likelihood}
The ELBO for $Y$ is derived by treating $Y$ as the observed variable and $X$ as the unobserved/latent variable. We then derive two expressions for the conditional distribution of $X$ given $Y = \y$, one from the measurement process and the other from the reconstruction process. The lower-bound is tight when these are close to each other.
\begin{description}[leftmargin=*]
\item[Measurement process:]
    Model parameters are generated by $X \sim \pi_X$ whereas the measured data are generated by the $\Y$-valued random variable $\left(Y|X=\x\right)\sim \mathcal{N}\left(\Op{A}(\x),\sigma_e^2\,\boldsymbol{I}\right)$ for given $\x \in \X$. 
    Let $\pi^{(m)}_{X,Y}$ denote the induced joint distribution of $(X,Y)$ with $\pi^{(m)}_{X}(\x):=\pi_X(\x)$ and $\pi^{(m)}_{Y}(\y)$ denoting its marginals.
    Also, let $\pi^{(m)}_{X|Y}(\x|\y)$ and $\pi^{(m)}_{Y|X}(\y|\x)$ denote the corresponding conditional distributions.
\item[Reconstruction process:]
    Data are generated by $Y \sim \pi_Y$ and reconstructed model parameters are generated by the $\X$-valued random variable $\left(X|Y=\y\right)\sim \mathcal{N}\left(\G_{\theta}(\y),\sigma_1^2\,\boldsymbol{I}\right)$ for given $\y \in \Y$. Denote the associated joint distribution by $\pi^{(r)}_{X,Y}(\x,\y)$ and its marginals and corresponding conditionals are denoted similarly as for the measurement process. 
\end{description}
% The $Y$-marginal from the measurement process depends on the model parameter $\theta$, whereas the $Y$-marginal from the reconstruction process corresponds to the true measurement distribution. Hence, applying a maximum likelihood (ML) principle amounts to maximizing 
The log-likelihood of $Y$ is given by $\mathcal{L}^{(y)}_{\text{ML}}(\theta)=\frac{1}{n_2}\sum_{i=1}^{n_2}\log \pi^{(m)}_Y(\y_i)$, which is the empirical average of the natural logarithm of the model-induced probability density computed over samples of the true distribution of $Y$.
% \begin{equation}
%     \mathcal{L}^{(y)}_{\text{ML}}(\theta)=\frac{1}{n_2}\sum_{i=1}^{n_2}\log \pi^{(m)}_Y(\y_i).
%     \label{mle_y_1st}
% \end{equation}
Using the statistical model above, $\mathcal{L}^{(y)}_{\text{ML}}(\theta)$ can be expressed as
\begin{align*}
\log \pi^{(m)}_Y(\y)
  &= \log\left(\int_{\X}\pi^{(m)}_{X,Y}(\x,\y)\,\mathrm{d}\x\right)=
\log\left(\mathbb{E}_{\pi^{(r)}_{X|Y}}\left[\frac{\pi^{(m)}_{X,Y}(X,\y)}{\pi^{(r)}_{X|Y}(X|\y)}\right]\right).
%\label{log_likeli_y}
\end{align*}
% \log\left(\int_{\X}\pi^{(m)}_{X,Y}(\x,\y)\frac{\pi^{(r)}_{X|Y}(\x|\y)}{\pi^{(r)}_{X|Y}(\x|\y)}\,\mathrm{d}\x\right)
% \nonumber\\
Since $\log$ is a concave function, applying Jensen's inequality leads to
\begin{align*}
\log \pi^{(m)}_Y(\y)
&\geq  \mathbb{E}_{\pi^{(r)}_{X|Y}}\left[\log\left(\frac{\pi_X(X)\pi^{(m)}_{Y|X}(\y|X)}{\pi^{(r)}_{X|Y}(X|\y)}\right)\right].
%\label{log_likeli_y_lb}
\end{align*}
% \mathbb{E}_{\pi^{(r)}_{X|Y}}\left[\log\left(\frac{\pi^{(m)}_{X,Y}(X,\y)}{\pi^{(r)}_{X|Y}(X|\y)}\right)\right]=
The above can further be simplified as
\begin{align}
\log \pi^{(m)}_Y(\y)
&\geq \mathbb{E}_{\pi^{(r)}_{X|Y}}\left[\log\left(\pi^{(m)}_{Y|X}(\y|X)\right)\right]+\mathbb{E}_{\pi^{(r)}_{X|Y}}\left[\log\left(\frac{\pi_X(X)}{\pi^{(r)}_{X|Y}(X|\y)}\right)\right]
\nonumber\\
&= \mathbb{E}_{\pi^{(r)}_{X|Y}}\left[\log\left(\pi^{(m)}_{Y|X}(\y|X)\right)\right]-\operatorname{KL}\left(\pi^{(r)}_{X|Y=\y},\pi_X\right).
\label{log_likeli_y_lb_simple}
\end{align}
Under the postulated statistical model, we have that
\begin{equation}
\pi^{(m)}_{Y|X}(\y|\x) := \mathcal{N}\left(\Op{A}(\x),\sigma_e^2\,\boldsymbol{I}\right)
\text{ and } 
\pi^{(r)}_{X|Y}(\x|\y):=\mathcal{N}\left(\G_{\theta}(\y),\sigma_1^2\,\boldsymbol{I}\right).
\label{cons_of_model}
\end{equation}
If the forward operator $\Op{A}$ is linear (which reduces to a matrix in the finite-dimensional case), then by  \eqref{cons_of_model} one can simplify the bound in \eqref{log_likeli_y_lb_simple}:
\begin{equation}
-\log \pi^{(m)}_Y(\y)\leq 
\operatorname{KL}\left(\pi^{(r)}_{X|Y=\y},\pi_X\right) + \frac{1}{2\,\sigma_e^2}\Bigl\|\y-\Op{A}\bigl(\G_{\theta}(\y)\bigr)\Bigr\|_2^2+ c_1.
\label{log_likeli_y_lb_final1}
\end{equation}
Here, $c_1$ is a constant independent of $\theta$ (see Proposition \ref{prop_gauss_expect} for a proof).

\subsubsection{Bound on the image likelihood}
The ELBO corresponding to $X$ can be derived by treating $X$ as the observed variable and the clean (synthetic) data $U=\Op{A}(X)$ as the latent variable. 
\begin{description}[leftmargin=*]
\item[Backward process:] 
    Here $U\sim \pi_{U}(\boldsymbol{u})$ and  $\left(X|U=\boldsymbol{u}\right)\sim \mathcal{N}\left(\G_{\theta}(\boldsymbol{u}),\sigma_2^2\,\boldsymbol{I}\right)$ for given $\boldsymbol{u}$, with possibly $\sigma_2\ll\sigma_1$.
\item[Forward process:] 
    $X\sim \pi_{X}$ and $\left(U|X=\x\right)\sim \delta\left(\boldsymbol{u}-\Op{A}(\x)\right)$ (Dirac measure concentrated at $\boldsymbol{u}=\Op{A}(\x)$).
\end{description}
Proceeding similarly to the analysis used for deriving a bound on the data likelihood, we can show that (with the superscripts $(f)$ and $(b)$ indicating the forward and backward processes, respectively)
\begin{equation}
\log \pi^{(b)}_X(\x)
  \geq \mathbb{E}_{U\sim \pi^{(f)}_{U|X=\x}}\left[\log\left(\pi^{(b)}_{X|U}(\x|U)\right)\right]-\underbrace{\operatorname{KL}\left(\pi^{(f)}_{U|X=\x},\pi_U\right)}_{\text{does not depend $\theta$}}.
\label{log_likeli_x_lb_simple}
\end{equation}
Using the postulated distributions to simplify the first term in \eqref{log_likeli_x_lb_simple} leads to 
\begin{equation}
-\log \pi^{(b)}_X(\x)\leq  \frac{1}{2\sigma_2^2}\Bigl\|\x-\G_{\theta}\bigl(\Op{A}(\x)\bigr)\Bigr\|_2^2+\operatorname{KL}\left(\pi^{(f)}_{U|X=\x},\pi_U\right).
\label{log_likeli_x_lb_final1}
\end{equation}
\subsubsection{Evidence bound}
The idea is now to combine \eqref{log_likeli_y_lb_final1} and \eqref{log_likeli_x_lb_final1}. 
Then, observe that minimizing the so-called (negative) evidence bound on the overall negative log-likelihood in \eqref{overall_ml} can be phrased as follows:
\begin{equation}
   \min_{\theta}\operatorname{KL}\left(\pi^{(r)}_{X|Y=\y},\pi_X\right) + \frac{1}{2\,\sigma_e^2}\Bigl\|\y-\Op{A}\bigl(\G_{\theta}(\y)\bigr)\Bigr\|_2^2+ \frac{1}{2\sigma_2^2}\Bigl\|\x-\G_{\theta}\bigl(\Op{A}(\x)\bigr)\Bigr\|_2^2.
    \label{overall_ml_final}
\end{equation}
This is identical to minimizing the ALPD training loss in \eqref{alpd_train_loss} but using the KL divergence instead of the Wasserstein-1  to quantify similarity in distribution.
\begin{prop}
\label{prop_gauss_expect}
Let $X\in\mathbb{R}^{d_x}$, $Y\in\mathbb{R}^{d_y}$, and let $\Op{A}\in \mathbb{R}^{d_y \times d_x}$ be a $d_y \times d_x$ matrix. Then, $\mathbb{E}_{X\sim \mathcal{N(\boldsymbol{\mu}, \boldsymbol{K})}}\left[\left\|Y-\Op{A}X\right\|_2^2\right]=\left\|Y-\Op{A}\boldsymbol{\mu}\right\|_2^2+\text{trace}\,(\Op{A}^\top \Op{A}\boldsymbol{K})$.
\end{prop}
\noindent\textbf{Proof}: Expanding the squared $\ell_2$-norm, we have that
\begin{eqnarray}
    \mathbb{E}\left[\left\|Y-\Op{A}X\right\|_2^2\right]&=&\mathbb{E}\left[Y^\top Y - 2\,Y^\top \Op{A}X+X^\top \Op{A}^\top \Op{A} X\right].
    \label{exp_quad_form_1}
\end{eqnarray}
Since the expectation is a linear operation, the expected value of the second term in \eqref{exp_quad_form_1} is $\left(-2\,Y^\top \Op{A}\boldsymbol{\mu}\right)$. The expected value of the third term can be evaluated as 
% \begin{multline}
%     \mathbb{E}\left[X^\top \Op{A}^\top \Op{A} X\right]= \mathbb{E}\left[\text{trace}\,(\Op{A}^\top \Op{A} XX^\top)\right]
%     \label{exp_quad_form_2}
% \end{multline}
\begin{eqnarray}
   \mathbb{E}\left[X^\top \Op{A}^\top \Op{A} X\right]&=& \mathbb{E}\left[\text{trace}\,(\Op{A}^\top \Op{A} XX^\top)\right] 
%   &=& \text{trace}\left(\Op{A}^\top \Op{A}\left(\mathbb{E}_{X\sim \mathcal{N(\boldsymbol{\mu}, \boldsymbol{K})}}\left[XX^\top\right]\right)\right)\nonumber\\
   =\text{trace}\,(\Op{A}^\top \Op{A}(\boldsymbol{K}+\boldsymbol{\mu}\boldsymbol{\mu}^\top))\nonumber\\
   &=&\text{trace}\,(\Op{A}^\top \Op{A}\boldsymbol{K})+\boldsymbol{\mu}^\top \Op{A}^\top \Op{A} \boldsymbol{\mu}.
    \label{exp_quad_form_2}
\end{eqnarray}
Substituting \eqref{exp_quad_form_2} in \eqref{exp_quad_form_1} leads to the desired result.\hfill $\blacksquare$
\begin{algorithm}[t]
\caption{Adversarial training of an iterative reconstruction network.}
\begin{algorithmic}
\STATE {\bf  1.} {\bf Input:} Gradient penalty $\lambda_{\text{gp}}$, initial reconstruction network parameter $\theta$ and critic parameter $\alpha$, batch-size $n_{b}$, Adam optimizer parameters $\left(\eta,\beta_1,\beta_2\right)$, the number of $\D$ updates per $\G$ update (denoted as $K$), penalty parameters $\lambda_{\X}$ and $\lambda_{\Y}$. 
\STATE {\bf  2.} {\bf for mini-batches $m=1,2,\cdots$, do (until convergence)}: 
\begin{itemize}[leftmargin=*]
\item Sample $\x_j\sim\pi_X$, $\y_j\sim \pi_Y$, and $\epsilon_j\sim \text{uniform}\,[0,1]$; for $1 \leq j \leq n_b$. Compute $\x^{(\epsilon)}_j=\epsilon_j \x_j+\left(1-\epsilon_j\right)\G_{\theta}(\y_j)$.
\item Critic loss: $\mathcal{L}_{\D}=\frac{1}{n_b}\sum_{j=1}^{n_b}\left[\D_{\alpha}(\G_{\theta}(\y_j))-\D_{\alpha}(\x_j) + \lambda_{\text{gp}}\left(\left\|\nabla \D_{\alpha}\left(\x^{(\epsilon)}_j\right)\right\|_2-1\right)_{+}^2\right]$.
% \begin{eqnarray}
%   \mathcal{L}_{\D}=\frac{1}{n_b}\sum_{j=1}^{n_b}\left[\D_{\alpha}(\G_{\theta}(\y_j))-\D_{\alpha}(\x_j) + \lambda_{\text{gp}}\left(\left\|\nabla \D_{\alpha}\left(\x^{\epsilon}_j\right)\right\|_2-1\right)_{+}^2\right].
%   \label{critic_loss_algo}
% \end{eqnarray}
\item \textbf{for $k=1,\cdots,K$, do}: update critic as $\alpha\leftarrow\text{Adam}_{\eta,\beta_1,\beta_2}\left(\alpha,\nabla_{\alpha}\mathcal{L}_{\D}\right)$.
% \begin{itemize}
%     \item 
% \end{itemize}
\item Compute the loss for the reconstruction network for the current mini-batch:
\begin{flalign*}
    \mathcal{L}_{\G}=\frac{1}{n_b}\sum_{j=1}^{n_b}\left[-\D_{\alpha}(\G_{\theta}(\x_j))+\lambda_{\X}\left\|\G_{\theta}\left(\Op{A}(\x_j)\right)-\x_j\right\|_2^2+\lambda_{\Y}\left\|\Op{A}\left(\G_{\theta}(\y_j)\right)-\y_j\right\|_2^2\right].
\end{flalign*}
\item Update reconstruction network parameters: $\theta\leftarrow\text{Adam}_{\eta,\beta_1,\beta_2}\left(\theta,\nabla_{\theta}\mathcal{L}_{\G}\right)$.
\end{itemize}
\STATE {\bf  3.} {\bf Output:} The trained iterative reconstruction network $\G_{\theta}$.
\end{algorithmic}
\label{algo_acr_train}
\end{algorithm}
%%%%%%%%%%%%%%%%%%%%%%%%

% In the learned primal-dual (LPD) approach \cite{lpd_tmi} proposed by Adler and \"Oktem, a parametric reconstruction operator inspired by the Chambolle-Pock algorithm \cite{pdhg_cp}, as explained in Sec.~\ref{g_phi_pdhg_sec}, is trained in a supervised manner to approximate the ground-truth from the CT projection data. We use essentially the same parametrization strategy as in \cite{lpd_tmi} and the one used for $\G_{\theta}$ in Sec.~\ref{g_phi_pdhg_sec}, however, unlike \cite{pdhg_cp}, we propose a new unsupervised training protocol that does not require paired samples. 

\subsection{Parametrizing the reconstruction and the critic networks}
\label{g_phi_pdhg_sec}
For parametrizing the reconstruction network $\G_{\theta}$, we adopt the same strategy as in \cite{lpd_tmi}, which is briefly explained here to make the exposition self-contained. The architecture of $\G_{\theta}$ is built upon the idea of \textit{iterative unrolling}, the origin of which can be traced back to the seminal work by Gregor and LeCun \cite{lecun_ista} on learned sparse approximation. Specifically, our reconstruction network $\G_{\theta}$ is parametrized by unrolling the Chambolle-Pock (CP) algorithm \cite{cham_pock} for non-smooth convex optimization. The CP algorithm is an iterative primal-dual scheme aimed at minimizing objectives of the form $f(\mathcal{K}\x)+g(\x)$, where $\mathcal{K}$ is a bounded linear operator, and $g$ and $f^*$ (the convex conjugate of $f$) are proper, convex, and lower semi-continuous. For convex $\R(\x)$ in \eqref{var_recon}, a wide range of problems are solvable by the CP algorithm, the update rules of which are given by
\begin{eqnarray}
    \boldsymbol{h}^{(\ell+1)} &=& \text{prox}_{\sigma\,f^*}(\boldsymbol{h}^{(\ell)}+\sigma\,\mathcal{K}(\bar{\x}^{(\ell)})),
    \boldsymbol{x}^{(\ell+1)} =\text{prox}_{\tau\,g}(\boldsymbol{x}^{(\ell)}-\tau\,\mathcal{K}^*(\boldsymbol{h}^{(\ell+1)})),\nonumber\\
    \bar{\x}^{(\ell+1)} &=& \x^{(\ell+1)}+\gamma (\x^{(\ell+1)}-\x^{(\ell)}), \text{\,\,for\,\,}0\leq\ell \leq L-1,
    \label{cp_update}
\end{eqnarray}
starting from a suitable initial point $\x^{(0)}=\bar{\x}^{(0)}$. In order to construct an architecture for $\G_{\theta}$, we essentially replace the proximal operators in \eqref{cp_update} by trainable convolutional neural networks (CNNs). More specifically, the output of $\G_{\theta}$ is computed by applying the following two steps repeatedly $L$ times:   
\begin{equation*}
    \h^{(\ell+1)}=\Gamma_{\theta_{\text{d}}^{(\ell)}}(\h^{(\ell)},\sigma^{(\ell)}\Op{A}(\x^{\ell}),\y), \text{\,and\,} \x^{(\ell+1)}=\Lambda_{\theta_{\text{p}}^{(\ell)}}(\x^{(\ell)},\tau^{(\ell)}\Op{A}^*(\h^{\ell+1})), 
\end{equation*}
The learnable parameters $\left\{\theta_{\text{p}}^{(\ell)}, \theta_{\text{d}}^{(\ell)},\sigma^{(\ell)},\tau^{(\ell)}\right\}_{\ell=0}^{L-1}$ are denoted using the shorthand notation $\theta$. The CNNs $\Gamma$ and $\Lambda$ are composed of a cascade of convolutional layers followed by a parametric ReLU activation. For the CT reconstruction experiment conducted in Sec. \ref{sec:numerical_expr}, we set $\boldsymbol{h}^{(0)}=\boldsymbol{0}$, and take the initial estimate $\x^{(0)}$ as the filtered back-projection (FBP) reconstruction. The number of layers is selected as $L=15$ and the filters in $\Gamma$ and $\Lambda$ are taken to be of size $5\times 5$ to increase the overall receptive field of the model to make it suitable for sparse-view CT. The critic $\D_{\alpha}$ (consisting of $\sim$ 2.76 million parameters) is a simple feed-forward CNN with four cascaded modules; each consisting of a convolutional layer, an instance-normalization layer, and a leaky-ReLU activation with negative-slope 0.2; followed by a global average-pooling layer in the end.
\begin{figure*}[h!]
	\centering
	% 1st row
	\begin{minipage}[t]{\textwidth}
		\centering
		\vspace{0pt}  
		\includegraphics[width=0.35\textwidth]{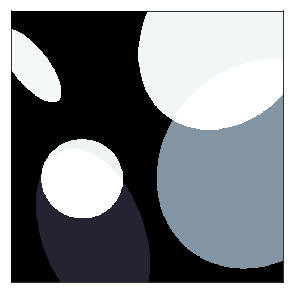}
		\qquad
		% trim=left bottom right top
		\includegraphics[width=0.35\textwidth,trim=30 27 0 0,clip,angle=90]{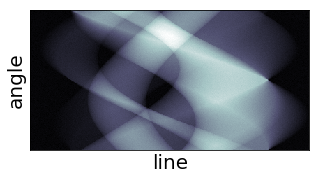}
		\vskip-0.1\baselineskip
		{\small Example of training data, image and its corresponding projection data}
	\end{minipage}	
	\\[1em] % 2nd row
	\begin{minipage}[t]{0.35\textwidth}
		\centering
		\vspace{0pt}  
		\includegraphics[width=\linewidth]{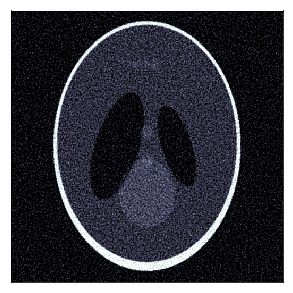}
		\vskip-0.5\baselineskip
		%{\scriptsize FBP: 19.5133 dB, SSIM 0.1286}
		{\small FBP: 19.51 dB, 0.13}
	\end{minipage}	
	\begin{minipage}[t]{0.35\textwidth}
		\centering
		\vspace{0pt}  
		\includegraphics[width=\linewidth]{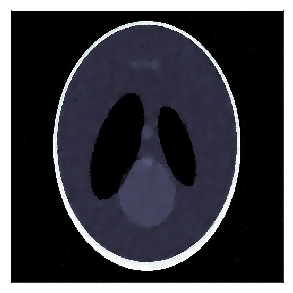}
		\vskip-0.5\baselineskip
		%{\scriptsize TV: 29.1795 dB, SSIM 0.8428}
		{\small TV: 29.18 dB, 0.84}
	\end{minipage}	\\
	\begin{minipage}[t]{0.35\textwidth}
		\centering
		\vspace{0pt}  
		\includegraphics[width=\linewidth]{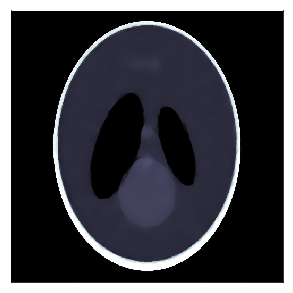}
		\vskip-0.5\baselineskip
		%{\scriptsize LPD: 27.8893 dB, SSIM 0.9558}
		{\small LPD: 27.89 dB, 0.96}
	\end{minipage}	
	\begin{minipage}[t]{0.35\textwidth}
		\centering
		\vspace{0pt}  
		\includegraphics[width=\linewidth]{expr_phantom/adv_lpd_5x5_filters_15layers_shepplogan_001}
		\vskip-0.5\baselineskip
		%{\scriptsize ALPD: 28.2668 dB, SSIM 0.8991}
		{\small ALPD: 28.27 dB, 0.90}
	\end{minipage}	
	\caption{\small{Comparison of supervised and unsupervised training on the Shepp-Logan phantom. The PSNR (dB) and SSIM are indicated below the images. ALPD does a better job of alleviating over-smoothing, unlike its supervised variant (LPD).}}
	\label{ct_image_figure_shepplogan}
\end{figure*}
% %%%%%%%%%%%%%%%%%%%%%%%%%%%%%%

\section{Numerical results}
\label{sec:numerical_expr}
For numerical evaluation of the proposed approach, we consider the classical inverse problem of sparse-view CT reconstruction. First, we demonstrate a proof-of-concept using phantoms containing random ellipses of different intensities for training the networks. Subsequently, we present a comparative study of the proposed ALPD approach with state-of-the-art model- and data-driven reconstruction methods. Parallel-beam projection data along 200 uniformly spaced angular directions, with 400 lines/angle, are simulated using the ODL library \cite{odl} with a GPU-accelerated \textit{astra} back-end. Subsequently, white Gaussian noise with a standard-deviation of $\sigma=2.0$ is added to the projection data to simulate noisy measurements. For supervised training, the phantoms and their corresponding noisy parallel-beam projections are aligned, whereas they are shuffled for unsupervised learning to eliminate the pairing information. 

The penalty parameters in \eqref{alpd_train_loss} are selected as $\lambda_{\X}=\lambda_{\Y}=10.0$, and the gradient penalty in Algorithm \ref{algo_acr_train} is also taken as $\lambda_{\text{gp}}=10.0$. The parameters in the Adam optimizer for updating both $\G_{\theta}$ and $\D_{\alpha}$ are chosen as $\left(\eta,\beta_1,\beta_2\right)=\left(5\times 10^{-5}, 0.50,0.99\right)$. The same set of hyper-parameters are used for training on both phantoms and real CT images. The critic $\D_{\alpha}$ is updated once per $\G_{\theta}$ update and the batch-size is taken as one (i.e., $K=1$ and $n_b=1$ in Algorithm \ref{algo_acr_train}).
%\vspace{-0.4in}
\subsection{Training on ellipse phantoms}
\label{phantom_train_sec}
%%%%
% \begin{figure*}[t]
% 	\centering
% 	\subfigure[\small{representative phantom}]{
% 		\includegraphics[width=2.20in]{expr_phantom/example_ellipse_phantom.png}}
% 	\subfigure[\small{corresponding projection data}]{
% 		\includegraphics[height=1.25in]{expr_phantom/example_ellipse_phantom_sinogram.png}}\\
% 	\subfigure[\small{FBP: 19.5133 dB, 0.1286}]{
% 		\includegraphics[width=2.20in]{expr_phantom/fbp_shepplogan_001.png}}
% 	\subfigure[\small{TV: 29.1795 dB, 0.8428}]{
% 		\includegraphics[height=2.20in]{expr_phantom/adv_lpd_5x5_filters_15layers_shepplogan_001.png}}\\
% 	\subfigure[\small{LPD: 27.8893 dB, 0.9558}]{
% 		\includegraphics[width=2.20in]{expr_phantom/lpd_5x5_filters_15layers_shepplogan_001.png}}
% 	\subfigure[\small{ALPD: 28.2668 dB, 0.8991}]{
% 		\includegraphics[height=2.20in]{expr_phantom/adv_lpd_5x5_filters_15layers_shepplogan_001.png}}	
% 		\caption{\small{Comparison of supervised and unsupervised training on the Shepp-Logan phantom. The proposed ALPD method does a better job of alleviating over-smoothing, unlike its supervised variant.}}
% 	\label{ct_image_figure_shepplogan}
% \end{figure*}
%%%%
%%%

In this experiment, we generate a set of 2000 2D phantoms, each of size $512\times 512$ and containing 5 ellipses of random eccentricities at random locations and orientations, for training the networks. Each ellipse has an intensity value chosen uniformly at random in the range [0.1, 1]. The intensity of the background is taken as 0.0 and the intensities of the ellipses add up in the regions where they overlap. A representative phantom and its corresponding noisy sparse-view parallel-beam projection are shown in the first row of Figure \ref{ct_image_figure_shepplogan}.

\indent The main objective of this experiment is to study the differences between supervised and unsupervised learning in terms of their ability to reproduce images containing homogeneous regions separated by sharp edges. For performance evaluation, we consider reconstruction of the Shepp-Logan phantom which essentially consists of elliptical homogeneous regions delineated by sharp boundaries. Since the total-variation (TV) regularizer, which seeks sparsity in the gradient image, is tailor-made for such phantoms, we consider the reconstructed image produced by TV as the `gold-standard' in this case and compare the proposed ALPD approach with its supervised counterpart vis-\`a-vis the TV reconstruction. To compute the TV solution, we use the ADMM-based solver in the ODL library with the penalty parameter $\lambda=10.0$, which leads to the best reconstruction in our setting.

\noindent A visual comparison of the reconstructed images using LPD and ALPD (in Figure \ref{ct_image_figure_shepplogan}) indicates that ALPD does a better job of recovering the three small tumors on the top region of the Shepp-Logan phantom. The ALPD reconstruction, although slightly inferior to TV in terms of PSNR, looks almost identical, while the supervised LPD reconstruction looks significantly blurry, making it difficult to discern the small tumors.    
\begin{figure*}[t]
% 1st row
\begin{minipage}[t]{0.32\textwidth}
  \centering
  \vspace{0pt}
  \begin{tikzpicture}[spy using outlines={circle,red,magnification=4.0,size=1.50cm, connect spies}]   
    \node {\includegraphics[width=\linewidth]{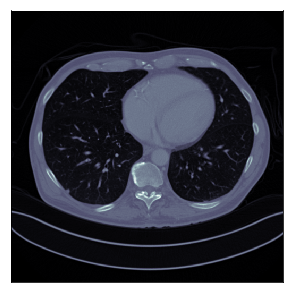}};
    \spy on (0.02,-0.68) in node [left] at (1.9,1.25);
    \spy on (-0.54,-0.76) in node [left] at (-0.4,1.25);  
  \end{tikzpicture}
  \vskip-0.5\baselineskip
  {\small Ground-truth}
\end{minipage}
\begin{minipage}[t]{0.32\textwidth}
  \centering
  \vspace{0pt}
  \begin{tikzpicture}[spy using outlines={circle,red,magnification=4.0,size=1.50cm, connect spies}]   
    \node {\includegraphics[width=\linewidth]{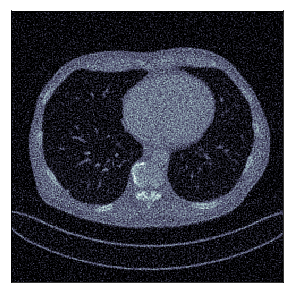}};    
    \spy on (0.02,-0.68) in node [left] at (1.9,1.25);
    \spy on (-0.54,-0.76) in node [left] at (-0.4,1.25);  
  \end{tikzpicture}
  \vskip-0.5\baselineskip  
  %{\small FBP: 21.6262 dB, 0.2435}
  {\small FBP: 21.63 dB, 0.24}
\end{minipage}
\begin{minipage}[t]{0.32\textwidth}
  \centering
  \vspace{0pt}
  \begin{tikzpicture}[spy using outlines={circle,red,magnification=4.0,size=1.50cm, connect spies}]   
    \node {\includegraphics[width=\linewidth]{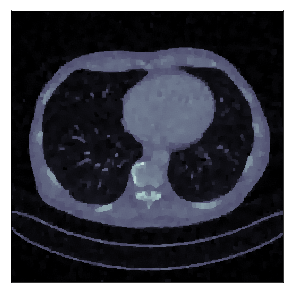}};
    \spy on (0.02,-0.68) in node [left] at (1.9,1.25);
    \spy on (-0.54,-0.76) in node [left] at (-0.4,1.25);  
  \end{tikzpicture}
  \vskip-0.5\baselineskip  
  %{\small TV: 29.2506 dB, 0.7905}
  {\small TV: 29.25 dB, 0.79}
\end{minipage}
\\[1em] % 2nd row
\begin{minipage}[t]{0.32\textwidth}
  \centering
  \vspace{0pt}
  \begin{tikzpicture}[spy using outlines={circle,red,magnification=4.0,size=1.50cm, connect spies}]   
    \node {\includegraphics[width=\linewidth]{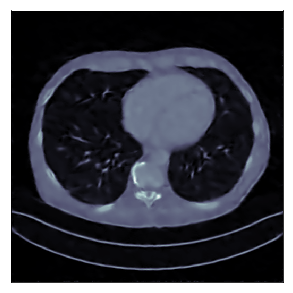}};
    \spy on (0.02,-0.68) in node [left] at (1.9,1.25);
    \spy on (-0.54,-0.76) in node [left] at (-0.4,1.25);  
  \end{tikzpicture}
  \vskip-0.5\baselineskip
  %{\small AR: 31.8257 dB, 0.8445}
  {\small AR: 31.83 dB, 0.84}
\end{minipage}
\begin{minipage}[t]{0.32\textwidth}
  \centering
  \vspace{0pt}
  \begin{tikzpicture}[spy using outlines={circle,red,magnification=4.0,size=1.50cm, connect spies}]   
    \node {\includegraphics[width=\linewidth]{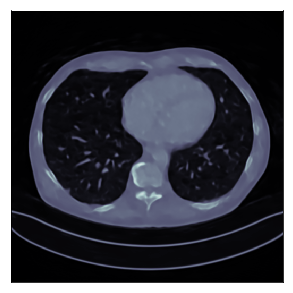}};    
    \spy on (0.02,-0.68) in node [left] at (1.9,1.25);
    \spy on (-0.54,-0.76) in node [left] at (-0.4,1.25);  
  \end{tikzpicture}
  \vskip-0.5\baselineskip  
  %{\small LPD: 33.3930 dB, 0.8824}
  {\small LPD: 33.39 dB, 0.88}
\end{minipage}
\begin{minipage}[t]{0.32\textwidth}
  \centering
  \vspace{0pt}
  \begin{tikzpicture}[spy using outlines={circle,red,magnification=4.0,size=1.50cm, connect spies}]   
    \node {\includegraphics[width=\linewidth]{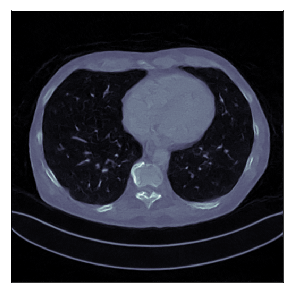}};
    \spy on (0.02,-0.68) in node [left] at (1.9,1.25);
    \spy on (-0.54,-0.76) in node [left] at (-0.4,1.25);  
  \end{tikzpicture}
  \vskip-0.5\baselineskip  
  %{\small ALPD: 32.4849 dB, 0.8419}
  {\small ALPD: 32.48 dB, 0.84}
\end{minipage}
\caption{\small{Comparison of ALPD with state-of-the-art model- and data-driven reconstruction methods on Mayo clinic data. The corresponding PSNR (dB) and SSIM are indicated below the images and the key differences in the reconstructed images are highlighted. The ALPD reconstruction is visibly sharper as compared to LPD, enabling easier identification of clinically important features.}}
\label{ct_image_figure_mayo2}
\end{figure*}
% %%%%%%%%%%%%%%%%%%%%%%%%%%

%%%%%%%%%%%%% Table by Shubo %%%%%%%%%
% \begin{table}[t]
%     \centering
%     %\setlength\tabcolsep{1.pt}
%     \begin{tabular}{l| c c r r r}
%     	\multicolumn{1}{c}{} 
%         &\multicolumn{1}{c}{\textbf{Method}} 
%         & \multicolumn{1}{c}{\textbf{ PSNR (dB) }} 
%         & \multicolumn{1}{c}{\textbf { SSIM }} 
%         & \multicolumn{1}{c}{\textbf{ \# param. }} 
%         & \multicolumn{1}{c}{\textbf{ Time (ms) }} \\
%         %\toprule
%         \hline
% 		\multirow{2}{*}{\textbf{analytical}}
%         & FBP & 21.2866  & 0.2043 & 1 & 14.0\\[1.2ex]
%         & TV & 30.3476  & 0.8110 & 1 & 21315.0 \\[1.2ex]
%         \hline
%         \multirow{2}{*}{\textbf{Supervised}}
%         %\midrule
%         & LPD & 35.1561  & 0.9048  & 854\,040 & 184.4 \\[1.2ex] 
%         %34.3590, 0.8889 with 3x3 filters, #params 411\,680; 35.1519 dB, 0.9073, 1138720 params with 5x5 filters and 20 layers 
%         & FBP + U-Net & 31.8008  &  0.7585 & 7\,215\,233 & 18.6 \\
%         \hline
%         \multirow{2}{*}{\textbf{Unsupervised}}
%         & AR & 33.6207  & 0.8750 & 19\,347\,890 & 41\,058.3\\[1.2ex]
%         & ALPD  & 33.7386  & 0.8559  &  854\,040 & 183.9\\[1.2ex]
%         %\bottomrule
%     \end{tabular}
%     \\[2ex]
%     \caption{\small{Average test performance for different reconstruction methods. ALPD has the same reconstruction time and number of parameters as LPD, but can be trained in an unsupervised manner.}}
%     \label{sparse_ct_table}
%\end{table}
%%%%%%%%%%%%%%%%%%%%%%%%%%%%%%%%%

% %%%%%%%%%%%%% Table by Ozan %%%%%%%%%
\begin{table}[t]
  \centering
  \begin{tabular}{l c l r r r}
        \multicolumn{1}{l}{\textbf{Method}} 
        & \multicolumn{1}{c}{\quad\textbf{PSNR (dB)}\quad} 
        & \multicolumn{1}{c}{\quad\textbf{SSIM}\quad} 
        & \multicolumn{1}{c}{\quad\textbf{\# param.}\quad} 
        & \multicolumn{1}{c}{\quad\textbf{Time (ms)}\quad} \\
        \toprule
        FBP & 21.2866  & 0.2043 & 1 & 14.0\\
        TV & 30.3476  & 0.8110 & 1 & 21\,315.0 \\%[1.2ex]
        \midrule
        \multicolumn{5}{l}{\emph{Trained against supervised data}} \\
        FBP + U-Net & 31.8008  &  0.7585 & 7\,215\,233 & 18.6 \\
        LPD & 35.1561  & 0.9048  & 854\,040 & 184.4 \\%[1.2ex] 
        %34.3590, 0.8889 with 3x3 filters, #params 411\,680; 35.1519 dB, 0.9073, 1138720 params with 5x5 filters and 20 layers 
        \midrule
        \multicolumn{5}{l}{\emph{Trained against unsupervised data}} \\
        AR & 33.6207  & 0.8750 & 19\,347\,890 & 41\,058.3\\
        ALPD  & 33.7386  & 0.8559  &  854\,040 & 183.9\\%[1.2ex]
        \bottomrule
  \end{tabular}
  \\[2ex]
  \caption{\small{Average performance of reconstruction methods. 
    LPD has the best overall performance (in terms of PSNR and SSIM), but this reconstruction method needs to be trained against 
    supervised data, i.e., pairs of high quality images (ground-truth) and corresponding noisy data. ALPD has slightly worse PSNR and SSIM values, but it can be trained against unsupervised data, which is vastly easier to get hold of as compared to supervised data. Note also that ALPD has significantly fewer parameters than AR, indicating that it can be trained against smaller datasets.}}
    \vspace{-0.2in}
  \label{sparse_ct_table}
\end{table}
% %%%%%%%%%%%%%%%%%%%%%%%%%%%%%%%%%
\subsection{Sparse-view CT on Mayo-Clinic data}
We perform a comparison of the proposed ALPD method with competing model- and data-driven reconstruction techniques on human abdominal CT scans released by the Mayo  Clinic  for  the  low-dose  CT  grand  challenge \cite{mayo_ct_challenge}. The dataset consists of CT scans corresponding to 10 patients, from which we extract 2D slices of size $512\times 512$ for our experiment. A total of 2250 slices extracted from 3D scans for 9 patients are used to train the networks in the data-driven methods, while 128 slices extracted from the scan for the remaining one patient are used for performance validation and comparison. The acquisition geometry and measurement noise distribution are kept the same as stated in Sec. \ref{phantom_train_sec}. For the sake of bench-marking the performance, we consider two model-based techniques, namely the classical FBP  and TV reconstruction. As two representative state-of-the-art data-driven methods, we consider adversarial regularization (AR) \cite{ar_nips}, and the LPD method \cite{lpd_tmi} trained on paired data. The performance of a U-Net-based learned post-processing applied on FBP is reported in Table \ref{sparse_ct_table} along with the aforementioned techniques as a baseline for fully data-driven methods.

% , but since it turns out to be considerably worse than LPD, we exclude it from the images shown in Figure \ref{ct_image_figure_mayo}. 
\indent Similar to what we noted for the Shepp-Logan phantom, the ALPD reconstruction outperforms LPD in terms of recovering sharp boundaries in the images, thus facilitating better delineation of clinically important features (see Fig. \ref{ct_image_figure_mayo2}). In terms of PSNR and SSIM, ALPD performs slightly worse than LPD, but it outperforms other competing techniques both qualitatively and quantitatively, as seen from the average PSNR and SSIM values reported in Table \ref{sparse_ct_table}. Notably, ALPD has the same reconstruction time as LPD, which is a couple of orders of magnitude lower than variational methods such as TV and AR that require computing iterative solutions to a high-dimensional optimization problem.   
\section{Conclusions}
We proposed an unsupervised training protocol that learns a parametric reconstruction operator for solving imaging inverse problems from samples of the marginal distributions of the image and the measurement. The reconstruction operator is parametrized by an unrolled iterative scheme, namely the Chambolle-Pock method, originally developed for solving non-smooth convex optimization in \cite{cham_pock} and subsequently adopted for network parametrization in the supervised learning framework in \cite{lpd_tmi}. The proposed learning strategy, nevertheless, is not limited to the specific parametrization of the reconstruction operator chosen in this work and extends, in principle, to other iterative reconstruction schemes. Experimental evidence suggests that the proposed method does not suffer from the curse of over-smoothing as it minimizes a distortion measure in the distribution space instead of seeking pixel-wise proximity. Minimizing the Wasserstein-1 distance requires the introduction of a critic network, leading to a more resource-intensive training, which pays off in terms of superior performance and a more flexible training framework that it offers. 

\bibliographystyle{splncs04}
\bibliography{ref}
\end{document}